\documentclass[]{spie}  %>>> use for US letter paper
\usepackage{amsmath,amsfonts,amssymb}
\usepackage{graphicx}
\usepackage[colorlinks=true, allcolors=blue]{hyperref}

\title{High Precision Astrometry Science in the Context of Space Mission Prospectives}

\author{Fabien Malbet}
\affil{Univ.\ Grenoble Alpes, CNRS, IPAG, 38000 Grenoble, France}
\author{Gary A.\ Mamon}
\affil{Sorbonne Université, CNRS, Institut d’Astrophysique de Paris, Paris, France}
\author{Lucas Labadie}
\affil{Univ.\ of  Cologne, Cologne, Germany}
\author{Alessandro Sozzetti}
\affil{Obs.\ Torino/INAF, Pino Torinese, Italy}
\author[1,5,6]{Manon Lizzana}
\affil{CNES, Toulouse, France}
\affil[6]{Pyxalis, Moirans, France}
\author[7]{Thierry Lépine}
\affil[7]{Institut d'Optique \& Hubert
  Curien Lab, Univ.\ de Lyon, Saint-Etienne, France}
\author[8]{Alain Léger}
\affil[8]{Univ.\ Paris-Saclay, CNRS, Institut d'astrophysique spatiale, Orsay, France}
\author[9]{Pierre-Olivier Lagage}
\affil[9]{Univ. Paris-Saclay, CEA, Saclay, France}

\authorinfo{Send correspondence to FM using the email address
  \texttt{Fabien.Malbet@univ-grenoble-alpes.fr}}

\begin{document} 
\maketitle

\begin{abstract}
  Astrometry is one of the oldest branches of astronomy which measures
  the position, the proper motion and parallax of celestial objects.
  Following the Hipparcos and Gaia missions that have measured several
  billions of them using global astrometry, we propose to increase
  astrometry precision on pointed objects using differential
  astrometry in a large field in order to unravel rocky planets in
  habitable zones of stars in the Sun vicinity and investigate the
  nature of dark matter in galactic environments as recommended by the
  ESA Senior Committee in the Voyager 2050 prospective. Substantial
  technology developments in a number of critical areas is needed in
  order to reach the highest required precision of
  sub-micro-arcsecond. One of them is CMOS image sensors using the
  stitching technique to merge the multiple design structures on the
  wafer and produce array with very large number of pixels. Another
  one is to calibrate the pixel positions using projecting modulating
  interferometric laser fringes on the array. Finally, the distortion
  of the optical system can be monitored and compensated using
  reference stars as metrology sources. The final precision depends on
  the diameter and the field of view of the telescope that is used as
  well as the time spent on each target. We present here the science
  goals that can be achieved with such missions either within the
  framework of an ESA Medium-class mission or even in the NASA most
  challenging Habitable Worlds Observatory, a large space telescope
  recommended by the American Astronomy and Astrophysics prospective
  for the 2020s and designed specifically to search for signs of life
  on planets orbiting other stars.
\end{abstract}

% Include a list of keywords after the abstract 

\section{Introduction}
\label{sec:intro}  % \label{} allows reference to this section

\begin{figure}[t]
  \centering
  \includegraphics[width=0.9\hsize]{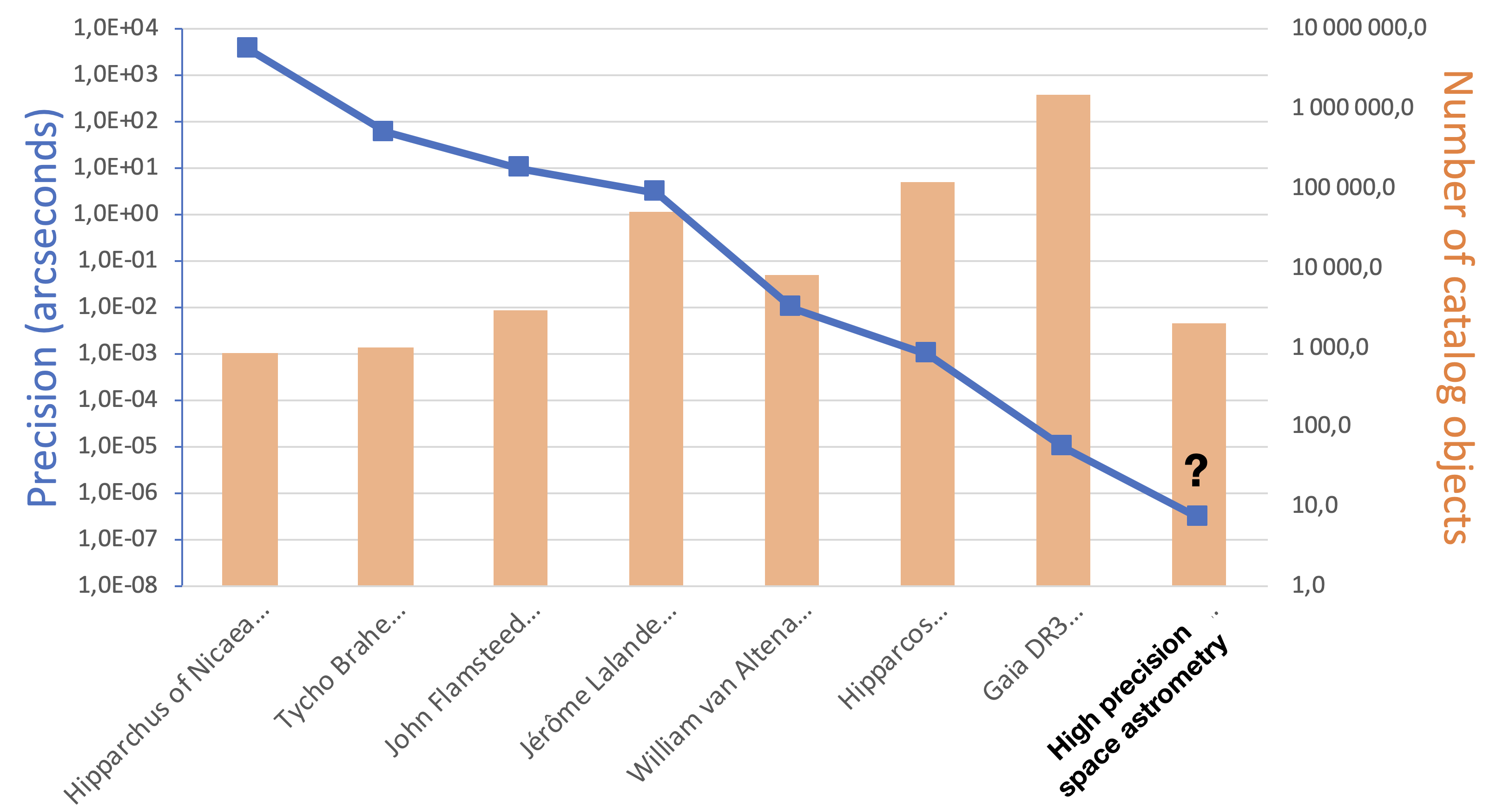}
  \caption{The progression of astrometric precision over time, from
    Hipparchus to Gaia and Theia}
  \label{fig:astrometry-ages}
\end{figure}

Astrometry, the science of measuring the precise positions and motions
of celestial objects, has been a foundational tool in astronomy for
centuries (Fig.~\ref{fig:astrometry-ages}). It began with the ancient
Greeks, with astronomers like Hipparchus of Nicaea cataloging
stars\cite{2024JHA....55..332G} as early as the second century BCE.
Over the centuries, observational precision advanced incrementally;
notable figures such as Tycho Brahe, John Flamsteed, and William van
Altena contributed to refining stellar position measurements. While
early astrometric efforts were limited to accuracies of several
arcminutes, the introduction of telescopes in the 17th century
revolutionized the field, making it possible to record stellar
positions with increasingly precise instruments. By the early 19th
century, telescopes equipped with advanced optics enabled measurements
accurate to within a few arcseconds, allowing astronomers to compute
stellar parallaxes and proper motions with greater confidence.

The launch of ESA Hipparcos mission in 1989 represented a
transformative moment in astrometry, as it achieved unprecedented
positional accuracy from space, free from Earth atmospheric
distortions. Hipparcos data enabled the creation of an extensive star
catalog\cite{1997ESASP1200.....E}, setting the stage for ESA Gaia
mission. Launched in 2013, Gaia brought global astrometry to a new
era, achieving several microarcsecond-level precision and mapping
several billion stars. This Gaia DR3 dataset\cite{2016A&A...595A...1G}
has allowed astronomers to probe the Milky Way structure in remarkable
detail, contributing to our understanding of stellar evolution,
galactic dynamics, and the formation history of the universe.  The
upcoming data releases from Gaia, DR4 and DR5, are expected to
incorporate meaurements of faint stellar companions down to the
planetary limit, with an anticipated discovery of approximately 10,000
Jupiter-mass planets. These releases promise to deepen our
understanding of the diversity of planetary bodies within our galaxy,
further reinforcing Gaia role in astrometric discovery.

Despite these advances, certain scientific challenges require even
higher precision and a more targeted approach. Both ESA \emph{Voyager
  2050}\cite{Voyage2050} prospective roadmap and the US National
Academies decadal survey for astronomy
\emph{Astro2020}\cite{2021pdaa.book.....N} underscore the need for
sub-microarcsecond precision to achieve critical scientific
objectives. Key priorities include the detection of Earth-mass
exoplanets orbiting nearby stars and the exploration of dark matter
influence within galactic environments. Gaia global astrometry
approach, while powerful, lacks the focused precision necessary for
detecting the wobble motions of nearby stars caused by orbiting
Earth-like planets. Similarly, understanding dark matter role in
shaping galactic structures calls for precise measurements of
positions and motions of faint stars that Gaia current capabilities
cannot achieve.

In response to these challenges, the Theia\footnote{In mythology,
  Theia is the daughter of Gaia and revered as the goddess of sharp
  vision, a fitting name for a mission designed to extend Gaia
  legacy with unparalleled precision in astrometric vision.} mission
proposes a new path forward. Designed as a medium-sized ESA mission,
Theia will achieve sub-microarcsecond precision using differential
astrometry, focusing on individual science targets with the aid of
reference stars in the field.  Theia high precision will be achieved
through innovative technologies, including a large CMOS detector array
with pixel-level calibration supported by interferometric
fringes. These advancements will provide the positional stability and
accuracy needed to reach Theia scientific goals.

\section{Science objectives }
\label{sec:mission-goals}

\begin{figure}[t]
  \centering
  \includegraphics[width=0.9\hsize]{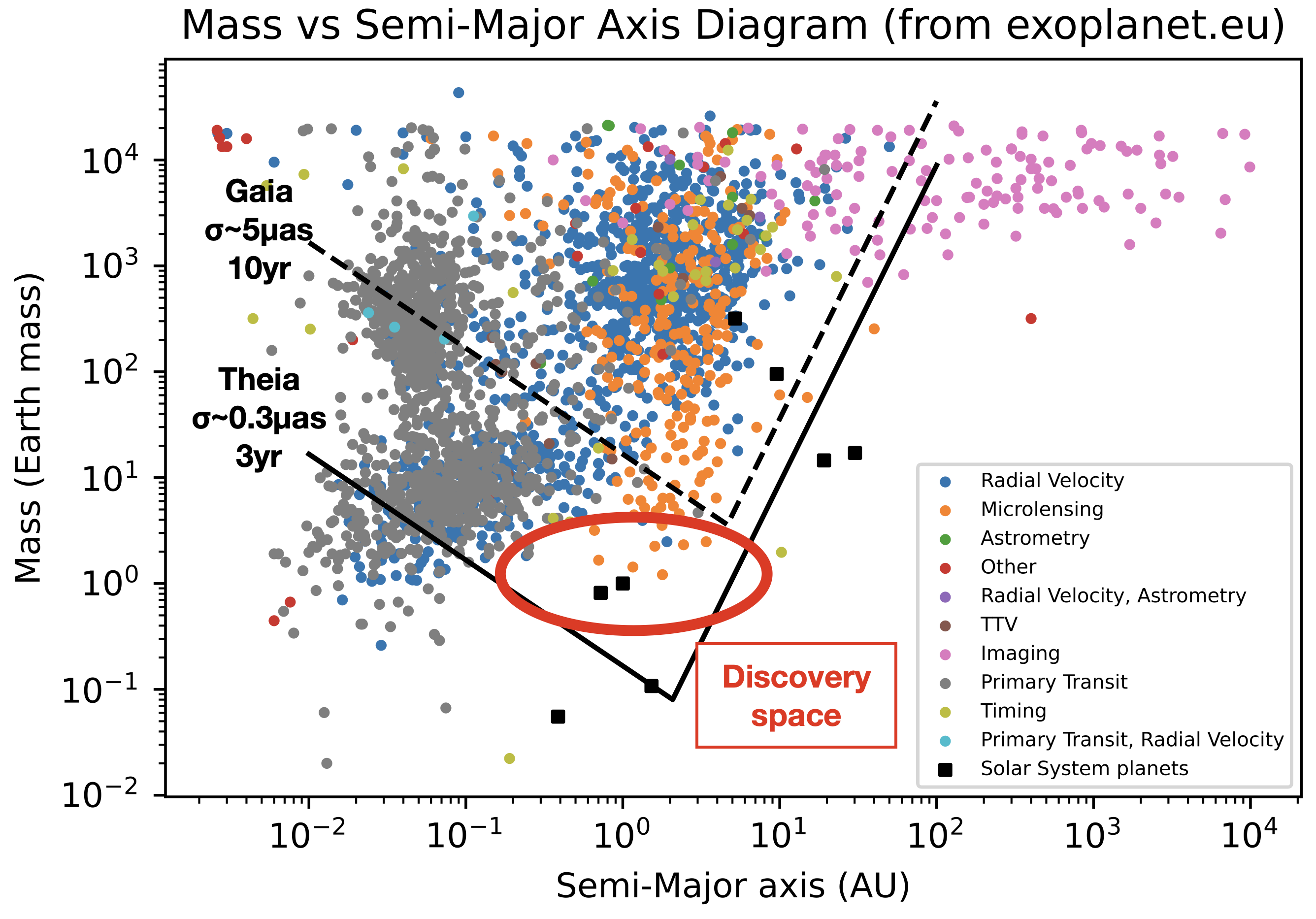}.
  \caption{Comparative plot for the discovery spaces of Gaia and
    Theia, emphasizing Theia higher precision which enables the
    exploration of a new discovery space (data extracted from the
    22-Oct-2024 version of the
    \href{https://exoplanet.eu}{Encyclopaedia of exoplanetary systems}
    catalog). }
  \label{fig:theia_discovery_space}
\end{figure}

The Theia mission has been designed\cite{2021ExA....51..845M} to
accurately track the positions of the stars over time in order to
fulfil two major scientific objectives.

The first main objective is the detection of rocky, Earth-like
exoplanets located in the habitable zones of nearby stars. By
employing differential astrometry, Theia can measure the subtle shifts
in a star position caused by the gravitational tug of orbiting
planets. This technique allows for the discovery of planets as small
as Earth within a distance of 15 parsecs, especially those orbiting
Sun-like stars. Such discoveries will refine the census of nearby
exoplanets and provide new insights into the formation and diversity
of planetary systems (Fig.~\ref{fig:theia_discovery_space}).

The second objective of Theia is to probe the nature of dark matter
(DM), in particular testing if it is collisionless and cold as assumed
in the usually assumed Cold Dark Matter (CDM) model. This can be done
by measuring stellar motions in specific galactic environments. By
studying dwarf spheroidal galaxies, where DM is thought to dominate
gravitational interactions at all scales, the proper motions of the
inner stars will test that the DM inner density profile is steep as
predicted in CDM for DM-dominated galaxies. Proper motions of stars
above and below the Milky Way disk may highlight the passage of the
numerous dark subhalos predicted by CDM. Finally, the directions of
proper motions of a few, distant, hypervelocity stars thought to
originate from 3-body encounters with the supermassive black hole at
the Galactic center will enable to test if the shape of the Milky
Way’s outer halo is prolate as predicted by CDM.

\section{Theia Mission profile}
\label{sec:mission-profile}

\begin{figure}[t]
  \centering
  \includegraphics[width=0.95\hsize]{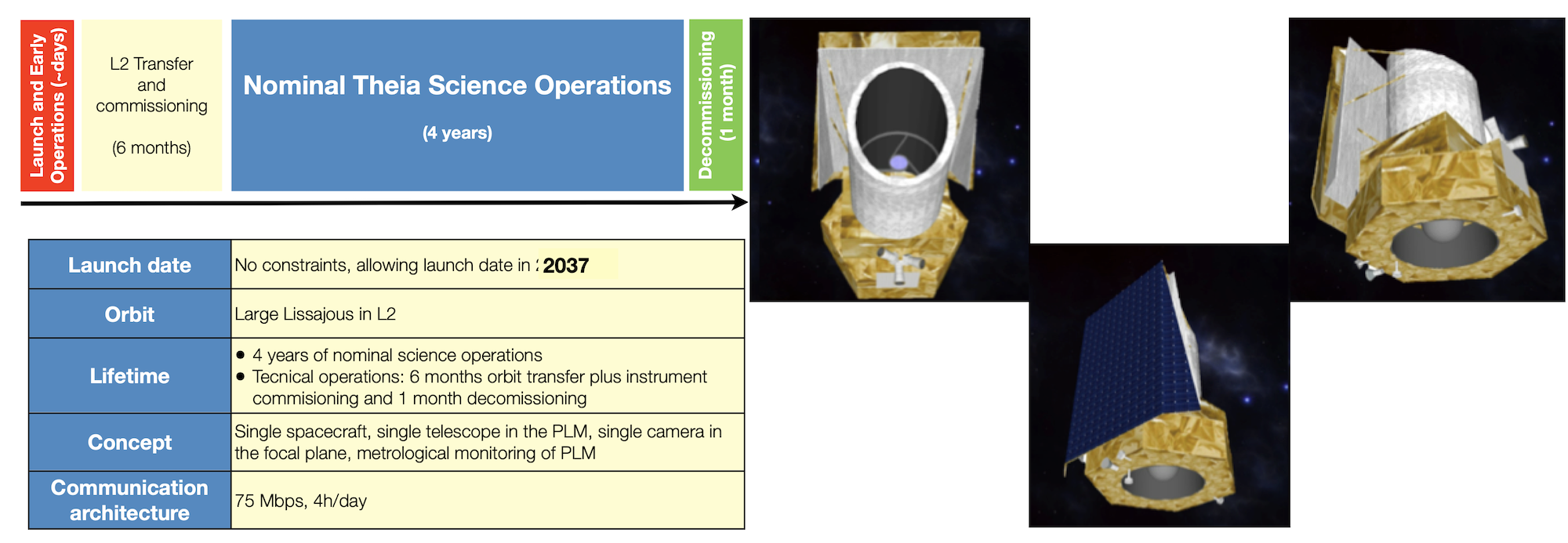}.
  \caption{Theia mission timeline and operational phases, including
    launch, transfer, and nominal science operations}
  \label{fig:mission-profile}
\end{figure}

The Theia mission (Fig.~\ref{fig:mission-profile}) will operate from a
Lissajous orbit around the Earth-Sun L2 point, providing a stable and
thermally consistent environment essential for long-term
high-precision measurements. The mission operational plan includes a
6-month transfer and commissioning phase, followed by four years of
nominal science operations and a one-month decommissioning period.

Theia telescope is designed as an 0.8-meter diffraction-limited
instrument with a 0.5-degree field of view, ideal for conducting
detailed astrometric observations of targeted fields. The telescope
optical design, a Korsch three-mirror anastigmat, minimizes
aberrations, supporting Theia high-precision objectives (see next session).

The telescope will operate in two modes to accommodate different
science cases: a “bright star mode” for detecting exoplanets around
nearby solar-type stars and a “faint observing mode” for observing
high-velocity stars and galactic structures. Each observation session
will be calibrated using metrology sources, ensuring consistency
across the mission lifetime.

\section{Theia optical design}
\label{sec:theia-optical-design}

A 3-mirror telescopes have much better corrected aberrations than 1
mirror (Newton) and 2-mirror (Cassegrain) telescopes. Among all
possible solutions for a 3-mirror telescope, we studied the Korsch
configuration\cite{1972ApOpt..11.2986K, 1977ApOpt..16.2074K,
  1980ApOpt..19.3640K, 2007SPIE.6687E..0SL} which is an anastigmat
(i.e.\ no spherical aberration, no coma, no astigmatism), with an almost
flat field, a raisonnable obstruction, and a large focal length that
should match the diffraction limit in the visible
($\lambda = 0.55\,\mu\mathrm{m}$) with the current size of CMOS
pixels.

\begin{figure}[t]
  \centering
  \includegraphics[width=0.9\hsize]{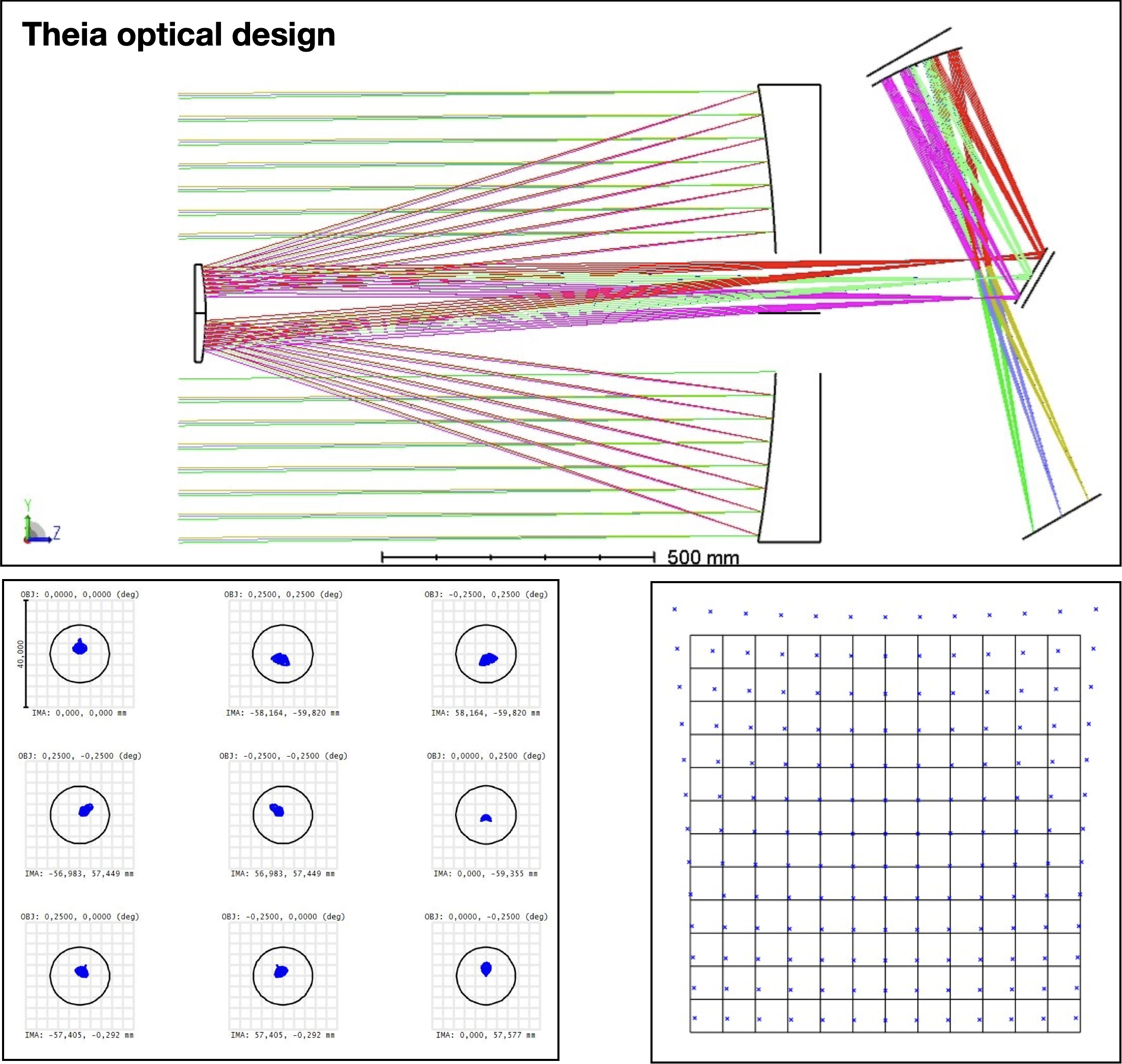}.
  \caption{Theia optical design using an anastigmat Korsch
    configuration with an almost flat field, a raisonnable
    obstruction, and a large focal length.}
  \label{fig:theia_optical_design}
\end{figure}

The design and the optimization of the telescope
(Fig.~\ref{fig:theia_optical_design}) were done using
\emph{Ansys-Zemax OpticStudio}.  The tolerancing and the stray light
will be studied later.  The specifications were the followings : the
primary mirror has a diameter of 80\,cm, the focal length is equal to
13 \,m, and the field of view is squared and equal to
$0.5^{\circ} \times 0.5^{\circ}$. Finally, the obstruction is governed
by the size of the mechanical mount of M2 and has a diameter of
24\,cm, ie. 30\% of the diameter of the primary.

We used for the degrees of freedom only the 3 radii, the 3 conic
constants and the back focal lenght. We used a field biais
($0.45^{\circ}$) in order to avoid the obstruction due to the flat
mirror between M2 and M3. But it is important to note that all the
mirrors are centered.  The solution we obtained on the top of
Fig.~\ref{fig:theia_optical_design} is diffraction limited, as can be
seen at the bottom left of Fig.~\ref{fig:theia_optical_design}. The
primary and tertiary mirrors are concave ellipsoids while the
secondary mirror is convex hyperboloid.  The distortion displayed at
the bottom righ is smaller than 2.2\%.

\section{Technological advancements in laboratory}
\label{sec:techn-advanc}

To achieve its ambitious science goals, Theia leverages recent
technological advancements in detector design, calibration, and data
processing. The mission success hinges on innovations that allow for
the precise detection of minute positional changes.

Theia imaging system use a CMOS detector array with stitched pixels,
enabling a large field of view while maintaining high positional
accuracy across the detector. This approach maximizes the effective
area of the sensor, essential for differential astrometry, and
minimizes the effect of pixel-level imperfections. Each pixel position
will be calibrated using interferometric laser
fringes\cite{pancher:hal-04752621} projected onto the array,
correcting for any distortions that could affect astrometric
measurements.

Additionally, Theia distortion correction system uses Gaia global
astrometric data to provide reference points for
calibration\cite{lizzana:hal-04746895}. This method significantly
reduces the telescope stability requirements, making it possible to
achieve high precision without the need for exceptionally stable
mechanical structures. With a focus on pointed differential
astrometry, Theia can accumulate many exposures for each target,
enhancing the accuracy of its measurements through repeated
observations.

To support Theia ambitious precision goals, the Institut de
Planétologie et d’Astrophysique de Grenoble (IPAG) has set up advanced
test benches\cite{pancher:hal-04752621} to refine and verify key
aspects of detector calibration and optical performance. These test
benches simulate Theia observational conditions, enabling 
testing of detector characterization, distortion correction, and
metrological accuracy. The setup includes an integrating sphere as a
stable light source, an OLED screen to simulate star fields, and a
high-precision GIGAPYX 4600 detector with a $4.4\mu\mathrm{m}$ pixel
pitch, which is used to test pixel-level response under various
lighting and distortion conditions. Each component is designed to ensure
that calibration measurements remain consistent with
the sub-microarcsecond accuracy required by Theia.

Furthermore, the test bench allows for assessments of field-dependent
distortions by emulating Theia operational field of view. This
enables the research team to fine-tune methods for real-time
distortion monitoring and correction, a critical capability for
maintaining high positional accuracy across the telescope wide
field. The insights gained from IPAG test benches will be instrumental
in refining Theia focal plane array design, ensuring that
laboratory-tested precision is maintained during in-flight operations.

More details on the technical work being performed in lab can be found
in this volume in Lizzana et al.\cite{lizzana:hal-04746895} and
Pancher et al\cite{pancher:hal-04752621}.

\section{Alternative with the HWO Wide Field Imager}
\label{sec:science-potent-impac}

In parallel with ESA high-precision astrometry prospectives, NASA
planned Habitable Worlds Observatory (HWO)\cite{2024SPIE13092E..1NF}
represents a significant step toward discovering exoplanets and
characterizing galacticenvironments at even greater observational
depths. The HWO, set to launch in the 2040s, is envisioned as a
large-aperture telescope operating across ultraviolet, visible, and
infrared wavelengths, with a primary mirror diameter of 6.5 to 8
meters. This instrument will allow for extremely detailed observations
of planetary systems around nearby stars, particularly targeting the
characterization of exoplanetary atmospheres within the habitable
zone.

The HWO observational design\cite{2024SPIE13092E..1NF} includes a
high-resolution imager with an anticipated point-spread function (PSF)
precision of 14 milliarcseconds (mas), which enables direct imaging
and astrometric analysis of exoplanets. Notably, for an Earth-like
planet located at 10 parsecs, HWO expected PSF can achieve an
astrometric signature detection down to 0.3 microarcseconds. This
level of precision, combined with a Nyquist-limited PSF resolution of
approximately $5. 10^{-5}$ pixels, positions HWO as a powerful
observatory for direct exoplanet characterization, though at a much
longer observational cadence compared to differential astrometric
missions.  The HWO will also feature a relatively modest field of view
of approximately $2 \times 3$ arcminutes, a limitation imposed by the
design for deep, focused investigations on individual targets.

Both Theia and HWO allows the limits of high-precision astrometry to
be pushed. In addition HWO direct imaging capabilities will provide
essential follow-up observations of exoplanets identified by
astrometric missions, particularly for spectroscopic characterization
of atmospheres. Theia differential astrometric measurements, on the
other hand, can be essential for identifying and cataloging habitable
exoplanets within 15 parsecs, providing high-accuracy stellar motion
data that can enhance subsequent direct imaging missions like HWO.

\section{Conclusion}
\label{sec:conclusion}

Theia represents a significant step forward in space-based astrometry,
building upon the legacy of Hipparcos and Gaia to achieve higher
precision and focused observational capabilities. By leveraging
cutting-edge technology in detector design and calibration, Theia will
open new frontiers in exoplanet discovery and dark matter research. If
successful, Theia will solidify Europe position in astrometric
innovation and contribute invaluable data to the global astronomical
community. Through its ambitious science objectives and technological
advancements, Theia promises to extend our understanding of the
universe composition and the possibility of life beyond Earth. The
Habitable Worlds Observatory (HWO) presents also an alternative path
for achieving high-precision space astrometry, offering a
complementary approach that warrants close consideration as we advance
toward the next generation of astrometric missions.

\begin{acknowledgements}
  The authors would like to acknowledge the
  contributions of the researchers and engineers who are not
  co-authors of this article but who have participated in the proposed
  missions and provided valuable input in response to ESA successive
  calls for proposals: NEAT (M3), micro-NEAT (S1), and Theia (M4,
  M5, M7).

  With regard to the funding of our research, we would like to
  acknowledge the support of the LabEx FOCUS ANR-11-LABX-0013 and the
  CNES agency. ML would like to acknowledge the support of her PhD
  grant from CNES and Pyxalis.
  
  This research has made use of NASA Astrophysics Data System
  Bibliographic Services.
\end{acknowledgements}

% References
\bibliography{malbet_icso2024} % bibliography data in report.bib

\begin{thebibliography}{10}

\bibitem{2024JHA....55..332G}
{Grasshoff}, G. and {Hoffmann}, S.~M., ``{An astronomical analysis of the data
  in the pseudo-Hipparchus palimpsest in the Codex Climaci Rescriptus},'' {\em
  Journal for the History of Astronomy}~{\bf 55},  332--349 (Aug. 2024).

\bibitem{1997ESASP1200.....E}
{Perryman}, M., {O’Flaherty}, K., {Heger}, D., and {McDonald}, A., eds.,
  [{\em {The HIPPARCOS and TYCHO catalogues. Astrometric and photometric star
  catalogues derived from the ESA HIPPARCOS Space Astrometry
  Mission}}{\nolinebreak\hspace{0.1em}]}, {\em ESA Special Publication} {\bf
  1200} (Jan. 1997).

\bibitem{2016A&A...595A...1G}
{Gaia Collaboration}, {Prusti}, T., {de Bruijne}, J.~H.~J., and et~al., ``{The
  Gaia mission},'' {\em \aap}~{\bf 595},  A1 (Nov. 2016).

\bibitem{Voyage2050}
{Tacconi}, L., {Arridge}, C., {Buonanno}, A., {Cruise}, M., {Grasset}, O.,
  {Helmi}, A., {Iess}, L., {Komatsu}, E., {Leconte}, J., {Leenaarts}, J.,
  {Martín-Pintado}, J., {Nakamura}, R., and {Watson}, D.,  [{\em {Final
  recommendations from the Voyage 2050 Senior
  Committee}}{\nolinebreak\hspace{0.1em}]}, ESA Special Publication, European
  Space Agency (2021).

\bibitem{2021pdaa.book.....N}
{National Academies of Sciences, Engineering and Medicine},  [{\em {Pathways to
  Discovery in Astronomy and Astrophysics for the
  2020s}}{\nolinebreak\hspace{0.1em}]}, National Academies Press (2021).

\bibitem{2021ExA....51..845M}
{Malbet}, F., {Boehm}, C., {Krone-Martins}, A., and et~al., ``{Faint objects in
  motion: the new frontier of high precision astrometry},'' {\em Experimental
  Astronomy}~{\bf 51},  845--886 (June 2021).

\bibitem{1972ApOpt..11.2986K}
{Korsch}, D., ``{Closed Form Solution for Three-Mirror Telescopes, Corrected
  for Spherical Aberration, Coma, Astigmatism, and Field Curvature},'' {\em
  \ao}~{\bf 11},  2986 (Dec. 1972).

\bibitem{1977ApOpt..16.2074K}
{Korsch}, D., ``{Anastigmatic three-mirror telescope.},'' {\em \ao}~{\bf 16},
  2074--2077 (Aug. 1977).

\bibitem{1980ApOpt..19.3640K}
{Korsch}, D., ``{Design and optimization technique for three-mirror
  telescopes},'' {\em \ao}~{\bf 19},  3640--3645 (Nov. 1980).

\bibitem{2007SPIE.6687E..0SL}
{Lampton}, M. and {Sholl}, M., ``{Comparison of on-axis three-mirror-anastigmat
  telescopes},'' in [{\em UV/Optical/IR Space Telescopes: Innovative
  Technologies and Concepts III}{\nolinebreak\hspace{0.1em}]},  {MacEwen},
  H.~A. and {Breckinridge}, J.~B., eds., {\em Society of Photo-Optical
  Instrumentation Engineers (SPIE) Conference Series} {\bf 6687},  66870S
  (Sept. 2007).

\bibitem{pancher:hal-04752621}
Pancher, F., Soler, S., Malbet, F., Lizzana, M., Kern, P., Lepine, T., and
  Leger, A., ``{Laboratory characterisation bench for high precision
  astrometry},'' in [{\em {International Conference on Space Optics (ICSO2024),
  Antibes, France, 21-25 October 2024}}{\nolinebreak\hspace{0.1em}]},
  Philippe, K., Fr{\'e}d{\'e}ric, B., Nikos, K., and Kyriaki, M., eds. (Oct.
  2024).
\newblock In this volume (preprint :
  \href{https://hal.science/hal-04752621}{hal-04752621}), in press.

\bibitem{lizzana:hal-04746895}
Lizzana, M., Malbet, F., Kern, P., Pancher, F., Soler, S., Lepine, T., and
  Leger, A., ``{Experimental tests of the calibration of high precision
  differential astrometry for exoplanets},'' in [{\em {International Conference
  on Space Optics (ICSO2024), Antibes, France, 21-25 October
  2024}}{\nolinebreak\hspace{0.1em}]},  Philippe, K., Fr{\'e}d{\'e}ric, B.,
  Nikos, K., and Kyriaki, M., eds. (Oct. 2024).
\newblock In this volume (preprint :
  \href{https://hal.science/hal-04746895}{hal-04746895}), in press.

\bibitem{2024SPIE13092E..1NF}
{Feinberg}, L., {Ziemer}, J., {Ansdell}, M., {Crooke}, J., {Dressing}, C.,
  {Mennesson}, B., {O'Meara}, J., {Pepper}, J., and {Roberge}, A., ``{The
  Habitable Worlds Observatory engineering view: status, plans, and
  opportunities},'' in [{\em Space Telescopes and Instrumentation 2024:
  Optical, Infrared, and Millimeter Wave}{\nolinebreak\hspace{0.1em}]},
  {Coyle}, L.~E., {Matsuura}, S., and {Perrin}, M.~D., eds., {\em Society of
  Photo-Optical Instrumentation Engineers (SPIE) Conference Series} {\bf
  13092},  130921N (Aug. 2024).

\end{thebibliography}
\bibliographystyle{spiebib} % makes bibtex use spiebib.bst

\end{document}